\documentstyle[12pt]{article}
\begin{document}

\begin{flushright}
{\sf Portsmouth University\\
Relativity and Cosmology Group \\
{\em Preprint} RCG} 95/9
\end{flushright}

\[ \]

\begin{center}
{\Large \bf Quadrupole anisotropy from photon quantum effects}
\[ \]
Alexei V. Nesteruk\footnote{School of Mathematical Studies, Portsmouth
University, PO1 2EG England} and
Roy Maartens${}^{1,}$\footnote{Member of Centre for Nonlinear Studies,
Witwatersrand University, South Africa}
\[ \]
{\bf Abstract}\\
\end{center}
\noindent
We consider quantum effects of an electromagnetic field in a
radiation--dominated
almost FRW spacetime. The dominant non--local quantum correction to
the photon distribution is a quadrupole moment, corresponding to an
effective anisotropic pressure in the energy--momentum tensor.

\[ \]
\section{Introduction}

Anisotropy in a cosmological photon distribution usually refers to the
cosmic background radiation after the decoupling of photons from
matter. Such anisotropy is predominantly caused by density
perturbations in the matter at the epoch of last scattering
\cite{cmb}. Here
we investigate another kind of anisotropy, not connected with density
perturbations or gravity wave perturbations, and not arising from
scalar field fluctuations in the inflationary epoch, but arising from
vacuum--polarisation type quantum effects of the electromagnetic
field itself.

We are only able to treat these effects in the case of
a field that is non--interacting with other matter fields. Therefore
the treatment is only really applicable to the radiation--dominated
epoch. These early--universe effects would
be washed out by the interaction of photons
with matter during the recombination epoch.
As such, our results are a contribution not to the theory of the
cosmic background radiation, but to quantum field theory in
curved spacetime and to early--universe dynamics.

We consider an early universe with small inhomogeneities and
anisotropies, i.e. with an almost Friedmann--Robertson--Walker
(FRW) geometry.
In order to obtain a stable vacuum state for massless fields and
to avoid infinite creation of particles in the vicinity of the
singularity $t=0$, we impose the Weyl curvature hypothesis of
Penrose \cite{penr79},
i.e. that the Weyl curvature should tend to zero as the singularity
is approached. Thus asymptotically, as $t\rightarrow 0$, the spacetime
becomes FRW. In order to invoke
standard results of quantum field theory, we further assume that the
spacetime is spatially homogeneous, so that the (small) deviations
from FRW geometry are only anisotropies. This is not unreasonable
for the early universe. Indeed, the anisotropies can be made as
small as desired, provided that they do not vanish exactly.
Non--vanishing anisotropy in the spacetime gives rise to the quantum
effects that produce effective anisotropy in the photon distribution
(quantum field theory is not analytic in the parameter describing the
degree of anisotropy).

Since we are primarily concerned with demonstrating
the existence of these effects
in principle, we will simplify the calculations by assuming
that the anisotropies are
of Bianchi I type, so that the metric has the form
\begin{equation}
ds^{2} =g_{\mu\nu}dx^\mu dx^\nu=
-dt^{2} + A_{i}^{2}(t) (dx^{i})^{2}
\label{metr}
\end{equation}
where $i=1,2,3$.
The scale factors $A_i$ must be such that the shear anisotropy is
$O(\epsilon)$, where $\epsilon$ is a smallness parameter:
\begin{equation}
H_i\equiv {\dot{A}_i\over A_i}=H+O(\epsilon),~~
H={\dot{A}\over A},~~A^3=A_1A_2A_3
\label{shear}
\end{equation}
The average
expansion is nearly that of a radiation--dominated FRW (spatially
flat) universe:
\begin{equation}
A\sim t^{1/2}+O(\epsilon)
\label{frw}
\end{equation}
As $t\rightarrow 0$, the Weyl curvature hypothesis requires that
$H_i\rightarrow H$ and $A\rightarrow at^{1/2}$ ($a$ constant).

The classical energy--momentum tensor is assumed to have the
perfect fluid radiation form to first order
\begin{equation}
T_{\mu\nu}={\textstyle{4\over3}}\rho u_\mu u_\nu+
{\textstyle{1\over3}}\rho g_{\mu\nu}
\label{em}
\end{equation}
where $u^\mu=\delta^\mu{}_0$ is the preferred four--velocity. This
energy--momentum tensor arises from the nearly equilibrium
classical photon distribution, which can be expressed in a
covariant harmonic expansion in terms of covariant
multipoles $F_{\mu\cdots}$ \cite{mes}:
\begin{equation}
f(t,k,e^\mu)=F(t,k)+F_\mu(t,k)e^\mu+F_{\mu\nu}(t,k)e^\mu e^\nu
+F_{\mu\nu\omega}(t,k)e^\mu e^\nu e^\omega+\cdots
\label{f}
\end{equation}
where $k$ is the wave number and $e^\mu$ is a unit vector orthogonal
to $u^\mu$. The monopole moment $F$ is Planckian to first order,
while the higher multipoles are $O(\epsilon)$. Only the first three
multipoles directly determine $T_{\mu\nu}$ \cite{mes}:
\begin{eqnarray}
\rho &=& \frac{4\pi}{A^4}\int_0^\infty F k^3 dk  \label{rho} \\
q_\mu &=& {4\pi\over 3 A^{4}}\int_0^\infty F_\mu k^3 dk  \label{q} \\
\pi_{\mu\nu} &=& {8\pi\over 15 A^{4}}\int_0^\infty F_{\mu\nu} k^3 dk
\label{pi}
\end{eqnarray}
The energy flux $q_\mu$ and anisotropic pressure $\pi_{\mu\nu}$
are zero (to first order) for the energy--momentum tensor
(\ref{em}). Although it is possible to achieve this with
non--zero $F_\mu$ and $F_{\mu\nu}$, it is less artificial and simpler
to take
\begin{equation}
F_\mu=0,~~F_{\mu\nu}=0
\label{f2}
\end{equation}
The third and higher multipoles, although $O(\epsilon)$, cannot
all vanish exactly, since this would force the shear anisotropy to
vanish and the geometry to reduce to FRW \cite{emt}.
With these conditions,
(\ref{f}) is a collision--dominated equilibrium distribution to
zero order. It satisfies the Boltzmann equation
\begin{equation}
p^\mu{\partial f\over\partial x^\mu}-\Gamma^\mu{}_{\nu\omega}p^\nu
p^\omega {\partial f\over\partial p^\mu}=C[f]=c_\mu e^\mu+\cdots
\label{bol}
\end{equation}
where $p^\mu$ is the four--momentum, and
the collision term $C[f]$ vanishes to zero order, so that
its multipoles $c_\mu,\cdots$ are $O(\epsilon)$.

\section{A specific model}

\subsection{Evolution of the universe}

To satisfy (\ref{shear}) and (\ref{frw}), we choose two times
$\tau, T$ such that the anisotropy is greatest during $\tau<t<T$.
This could correspond to a period where one species
decouples from the thermalised radiation.
Alternatively, we could postulate that this anisotropy
interval is imprinted by initial conditions at the singularity.
In any case, we choose an evolution of the scale factors
according to the ansatz
\begin{equation}
A_i(t) = a t^{1/2}
\left[ {1 + \left({t\over\tau}\right)^q \over
1+\left({t\over T}\right)^q}\right]^{\frac{\alpha_i-1/2}{q}}
\label{anzatz}
\end{equation}
where $q> 1$ and $|\alpha_i-{1\over2}|\ll 1$. Then
\[
A = a t^{1/2}
\left[ {1 + \left({t\over\tau}\right)^q \over
1+\left({t\over T}\right)^q}\right]^{\frac{\alpha-3/2}{3q}}
\]
where $\alpha\equiv \alpha_1+\alpha_2+\alpha_3$, so that
$|\alpha-{3\over2}|\ll 1$. If we choose $\alpha={3\over2}$,
then $A=at^{1/2}$ and the average scale factor and
average rate of expansion are exactly those of FRW.

The rates of expansion given by (\ref{anzatz}) are
\begin{eqnarray}
H_i &=& {1\over 2t}+(\alpha_i-{\textstyle{1\over2}})h(t) \label{hi}\\
H &=& {1\over 2t}+(\alpha-{\textstyle{3\over2}})h(t) \label{hub}
\end{eqnarray}
where
\begin{equation}
h(t) = t^{q-1} \left[
 \frac{T^{q} - \tau^{q}}{(t^{q} + \tau^{q})(t^{q} + T^{q})}\right]
\label{h}
\end{equation}

For simplicity, we assume axisymmetry $$A_2=A_1$$ and take $$\alpha
={3\over2}$$ Define $$\epsilon=\alpha_3-\alpha_1$$
as the anisotropy parameter. Then the shear anisotropy may be
measured by
\begin{equation}
\Delta H(t)\equiv H_3-H_1=\epsilon h(t)
\label{deltah}
\end{equation}
Thus $\Delta H\sim t^{q-1}$ for $t\ll\tau$, while
$\Delta H\sim t^{-(q+1)}$ for $t\gg T$. The anisotropy is strongly
suppressed towards the singularity and after the anisotropy
interval.

The behaviour of the Weyl curvature is given by
\begin{equation}
C_{\mu\nu\omega\sigma}\,C^{\mu\nu\omega\sigma} = \frac{1}{72}
\left( \frac {d}{dt} \Delta H + H \,  \Delta H \right)^2 \sim
t^{2(q-2)} \qquad \mbox{as} \, t \rightarrow 0
\label{WCH}
\end{equation}
on using (\ref{hub}), (\ref{deltah}) and the results of \cite{birdav}.
Thus the Weyl curvature hypothesis requires $q > 2$ \cite{nest94}.
It is shown in \cite{nest94}
that only for $q>2$ is it
possible to choose a well--defined initial vacuum state (i.e.
without particles) for matter/ radiation fields at
the cosmological singularity $t=0$.
Therefore the damping of shear anisotropy
near the singularity, which is consistent with the
(strong) Weyl curvature hypothesis, allows one to
choose the initial vacuum state, and to prevent instability of the
vacuum for massless conformal fields and infinite creation of
particles in the vicinity of the singularity. The
expansion law (\ref{anzatz}) thus makes the
singularity stable in relation to production of conformal
massless particles. (In fact it is also stable for
non--conformal particle creation \cite{nest94}.)
Henceforth we take $$q=3$$

\subsection{Quantum field and vacuum definition}

As a model of matter we consider an electromagnetic field in
the metric (\ref{metr}). The classical theory
is described in \cite{sagnotti81}. The field equations
and secondary quantization were studied in \cite{nest83}.
The time-dependent part of both transversal components of the
electromagnetic field with $A_{2} =A_{1}$ satisfies \cite{nest83}
\begin{equation}
{d^2\over d\eta^2} Y (\eta,k)+[{\textstyle{2\over 3}}
A\Delta H]{d\over d\eta} Y (\eta,k) + k^2 \delta^{2}Y (\eta,k) = 0
\label{Y}
\end{equation}
where $\eta$ is conformal time ($dt=Ad\eta$) and
\[
\delta^{2} = \sin^{2} \theta  \left(\frac{A_{3}}{A_{1}}\right)^{2/3}
          +   \cos^{2} \theta  \left(\frac{A_{1}}{A_{3}}\right)^{4/3}
\]
so that by (\ref{anzatz})
\begin{equation}
\delta=1+\epsilon \delta_{(1)}+O(\epsilon^2),~~\delta_{(1)}=
{\textstyle{1\over3}}(1-3\cos^2\theta)\log\left[{1+\left({t\over\tau}
\right)^3 \over 1+\left({t\over T}\right)^3}\right]
\label{del}
\end{equation}
If $\beta$ satisfies
\[
{d\beta\over d\eta} + ({\textstyle{1\over3}}A\Delta H)\beta = 0
\]
then by (\ref{Y}), the rescaled field $y=Y/\beta$ satisfies
\begin{equation}
{d^2y\over d\eta^2} + \left( k^2 \delta^{2} + m_{ef\!f}^{2}
\right) y = 0
\label{rescale}
\end{equation}
where the effective mass of the electromagnetic field is
\begin{equation}
m_{ef\!f}^2 =
- {d \over d\eta}({\textstyle{1\over3}}A\Delta H)
- ({\textstyle{1\over3}}A\Delta H )^{2}
\label{meff}
\end{equation}

To define the choice of the vacuum state, we introduce the
representation
\begin{equation} Y  = \left(\frac{k}{\delta}\right)^{1/2}
\left(\frac{A_{1}}{A_{3}}\right) (\Phi \, e_{-} + \Psi \, e_{+} ),
\qquad
\frac{dY}{d\eta} =
 i k \delta \left(\frac{k}{\delta}\right)^{1/2}
 \left(\frac{A_{1}}{A_{3}}\right) (\Phi \, e_{-} - \Psi \, e_{+}  )
\end{equation}
where
\[
e_\pm = \exp\left(\pm i\int k\delta d\eta\right)  = \exp \left( \pm i
\int {k \delta\over A}dt \right)
\]
and $\Phi, \Psi$ satisfy
\begin{equation}
{d \Phi \over d\eta} =  {\textstyle{1\over2}}W\Psi e_{+}^{2},\qquad
{d \Psi \over {d\eta}} ={\textstyle{1\over2}}W \Phi e_{-}^{2}
\label{(4.4b)}
\end{equation}
where $ W \equiv \Delta H\sin^2 \theta $.

It is natural to define the vacuum state at $\eta=\eta_0$ by demanding
that the positive frequency solutions be those which satisfy
$(d Y /d\eta)_{\eta_0} = i kY (\eta_0)$, or equivalently
$\Phi(\eta_0) = 1$ and $\Psi(\eta_0) = 0$.
(This is possible since (\ref{del}) shows that $\delta (\eta_0) = 1$.)
In the special case of an FRW universe with
$\Delta H = 0$ we have $m_{ef\!f}^2 = 0$
and $W = 0$ so  $\Phi =$const, $\Psi =$const, and the vacuum
defined in this way will be stable.  Thus one can use the isotropic
universe as an absolute standard for the number of photons produced
in an anisotropic universe. Correspondingly, such a universe would
be an absolute standard for the gravitational entropy which is
contained in the gravitational field.

\section{The energy-momentum tensor}
\subsection{Vacuum tensor}

To determine the quantum corrections to the components of
the energy--momentum tensor produced by
the expansion of our model universe, we adopt the approach
of Hu and Parker \cite{hupark77}. For those modes of the field where
we can not neglect
$ m_{ef\!f}$ we find the effect of parametric amplification and
instability of the vacuum accompanied by the creation or annihilation
of particles.  For modes of the field with $k \gg m_{ef\!f}A$, we can
neglect the effective mass and they satisfy equations which have
effectively isotropic form. For these modes there will be no
quantum particle creation. At the early times when photon
production is significant, $m_{ef\!f} \sim 1/t$, and the critical
momentum may be taken as $k_H(t) \equiv A(t)/t$, corresponding to
wavelengths of the order of the size of the particle horizon.

We separate the energy--momentum tensor of the electromagnetic
field into a vacuum polarisation term and a term
arising from higher frequency modes which can be treated classically.
It is these latter which  can be thought of as `real' photons.
They correspond to modes which were
parametrically amplified when their
wavelength was of the order of the particle horizon or longer $\bigl(k
\leq k_H(t)\bigr)$, when quantum effects were significant to them, but
which have since passed within the particle horizon and decoupled from
the background thereby forming a classical remnant.

Under the sudden approximation valid for such long wavelength modes,
the energy--momentum tensor of photons created on  scales longer than
the particle horizon and then flowing across it and accumulating
inside can be written as (compare with \cite{hupark77,nest91})
\begin{equation}
T_{\mu \nu}^C(t) = \int_{0}^t
                      \left[ {A(t')\over A(t)}\right]^{\!4}
     {\frac{\partial}{\partial k_H}} T_{\mu \nu}^Q(k_H(t'), t')
     \left[ -{d k_H(t') \over dt'}\right] d t'
\label{energ}
\end{equation}
Notice that we use the approximation that no particles are
created with $k > k_H$, and in (\ref{energ}) we do
not include the contribution from all particles created at time
$t'$, but only those that enter the horizon, because we only take
the partial derivative of $n_Q$ and $\rho_Q$ with respect to
$k_H$. In (\ref{energ}), $T_{\mu\nu}^Q$ is the vacuum expectation
value of the energy--momentum tensor operator for the massless field:
\[
T_{\mu\nu}^Q(t) = \langle 0_{t_{in}} \mid N_t
 \widehat T_{\mu\nu}(t,\vec x) \mid 0_{t_{in}}\rangle, \qquad
  N_{t} \widehat T_{\mu\nu} = \widehat T_{\mu\nu}
  - \langle 0_t\mid \widehat T_{\mu\nu}\mid 0_t\rangle,
\]
where $N_{t}$ indicates normal ordering at time $t$.
Here, the initial vacuum state $\vert 0_{t=0}\rangle$
is determined in the previous section and the
transition to the vacuum state $\vert 0_t\rangle$ at time $t$ is
done as usual by Bogoliubov transformations.

Finally we obtain (see also \cite{nest91,afld})
\begin{equation}
T_{\mu\nu}^C = {\frac{1}{(2\pi)^{2} A^{4}(t)}}
               \int_{0}^{t}  \, \tau_{\mu\nu}(t') \,
k_H^{3}(t') \left[ -{d k_H(t') \over d t'}\right] d t'
\label{energx}
\end{equation}
where
\[
\tau_{\nu}^{\mu}(t)  = \int  {\bar \tau}_{\nu}^{\mu} (t, \theta)
\sin\theta d\theta
\]
and the non-zero components of ${\bar \tau}_{\nu}^{\mu}$ are
\begin{eqnarray}
{\bar \tau}_0^0 &=&  4S\delta \nonumber\\
{\bar \tau}_1^1 &=&
{\bar \tau}_2^2 = - (2 S + U ) \delta\sin^2\theta \nonumber\\
{\bar \tau}_3^3 &=&  2
\left(- 2 S  \cos^2 \theta + U   \sin^2 \theta \right)\delta \nonumber
\end{eqnarray}
where
\[
S = | \Psi |^2 ,  \;    U =  2 {\rm Re} (\Psi^{*} \Phi e_+^2)
\]
In accordance with the conditions of applicability of the formula
(\ref{energx}), we can replace the exponentials $e_\pm$ by
1 in
(\ref{(4.4b)}) \cite{hupark77}. Taking the natural initial
conditions $\Phi = 1$ and $\Psi = 0$, one can obtain
$\Psi = \sinh R$, $\Phi = \cosh R$
where
\begin{equation}
R = \epsilon r,~~r(t,\theta)\equiv
{\textstyle{1\over2}}\sin^{2} \theta \int_{0}^{t}  h(t') dt'
\label{r}
\end{equation}
Together with (\ref{del}), this leads to an
expansion in $\epsilon$:
\[
\bar{\tau}^{\mu\nu}=\bar{\tau}_{(0)}^{\mu\nu}+\epsilon
\bar{\tau}_{(1)}^{\mu\nu}+\cdots
\]
where
\begin{eqnarray}
\bar{\tau}_0^0 &=& \left( 4r^2\right)\epsilon^2+O(\epsilon^3)
\label{00}\\
\bar{\tau}_1^1 &=& \bar{\tau}_2^2=-\left(2r\sin^2\theta\right)\epsilon
-\left(2r[r+\delta_{(1)}]\sin^2\theta\right)\epsilon^2+O(\epsilon^3)
\label{11}\\
\bar{\tau}_3^3 &=& \left(4r\sin^2\theta\right)\epsilon+
\left(4r[\delta_{(1)}\sin^2\theta-r\cos^2\theta]\right)\epsilon^2
+O(\epsilon^3) \label{33}
\end{eqnarray}

\subsection{Corrections to the classical tensor}
By (\ref{00}), (\ref{11}), (\ref{33}) and (\ref{energx}) it follows
that to first order, quantum corrections to the
energy--momentum tensor (\ref{em}) contribute only a trace--free
anisotropic pressure tensor:
\begin{equation}
T^T_{\mu\nu}=T_{\mu\nu}+T^C_{\mu\nu}={\textstyle{4\over3}}\rho u_\mu
u_\nu+{\textstyle{1\over3}}\rho g_{\mu\nu}+\epsilon\pi_{(1)\mu\nu}
+O(\epsilon^2)
\label{corremt}
\end{equation}
where
\begin{equation}
\pi_{(1)\mu\nu}=Q(t) a_{\mu\nu},~~a_\mu^\nu\equiv {\rm diag}
(0,1,1,-2)
\label{pi1}
\end{equation}
and $Q$ is determined by (\ref{energx}) with (\ref{r}), (\ref{11}).

The energy density is only corrected at second order, and no energy
flux occurs at any order, owing to the symmetries of (\ref{metr}).
By (\ref{pi}), it follows that the quantum corrections constitute
a quadrupole moment contribution $F^C_{\mu\nu}$ to the photon
distribution, where
\begin{equation}
\pi_{(1)\mu\nu}={8\pi\over 15A^{4}}\int_0^\infty F^C_{(1)\mu\nu}k^3dk
\label{quad}
\end{equation}
This gives rise to a quadrupole anisotropy in the temperature
fluctuation of radiation. Using the covariant formalism of \cite{mes},
we find
\begin{equation}
{\delta T\over T}(\vec{e})=\left({15Q a_{\mu\nu}e^\mu e^\nu \over
8\rho}\right)\epsilon
\label{temp}
\end{equation}
(Octopole and higher contributions at first order arise from the
classical multipoles $F_{\mu\nu\omega},\cdots$.)

The dynamics of the quadrupole anisotropy are governed by the
Boltzmann equation (\ref{bol}). The covariant monopole, dipole
and quadrupole moments of (\ref{bol}) for the distribution
(\ref{f}) (with quantum correction) and the spatially homogeneous
metric (\ref{metr}), follow from the general results
of \cite{emt}. If we linearise these Boltzmann multipoles,
we obtain (compare \cite{mes}):
\begin{eqnarray}
k\dot{F}-Hk^2{\partial F\over\partial k} &\approx & 0 \nonumber\\
0 &\approx & c_\mu \nonumber\\
k\dot{F}^C_{\mu\nu}-Hk^2{\partial F^C_{\mu\nu}\over\partial k}
-\sigma_{\mu\nu}k^2{\partial F\over\partial k} &\approx & c_{\mu\nu}
\label{qbol}
\end{eqnarray}
where $\sigma_{\mu\nu}$ is the shear anisotropy tensor:
\begin{equation}
\sigma_{\mu\nu}=-{\textstyle{1\over3}}\epsilon h a_{\mu\nu}
\label{sigma}
\end{equation}

The first (monopole) and second (dipole) equations
imply the linearised
energy--momentum conservation equations \cite{mes}. The first is
satisfied identically (to first order) since $F$ is nearly Planckian,
and the second because of spatial homogeneity, which also implies
that the dipole of the collision term vanishes to first order.
The third (quadrupole) equation (\ref{qbol}), when integrated, gives,
using (\ref{rho}) and (\ref{pi}),
an evolution equation for $\pi^C_{\mu\nu}$
(compare \cite{mes}):
\begin{equation}
\dot{\pi}^C_{\mu\nu}+4H\pi^C_{\mu\nu}\approx -{\textstyle{8\over15}}
\rho\sigma_{\mu\nu}+\gamma_{\mu\nu}
\label{pidot}
\end{equation}
where
\[
\gamma_{\mu\nu}={8\pi\over 15A^4}\int_0^\infty c_{\mu\nu}k^2dk
\]
is an average quadrupole collision term. By (\ref{pi1}) and
(\ref{sigma}), $\gamma_{\mu\nu}=\gamma(t) a_{\mu\nu}$, and
(\ref{pidot}) becomes
\begin{equation}
\dot{Q}+4HQ={\textstyle{8\over 45}}\rho h +\gamma
\label{qdot}
\end{equation}
Using (\ref{hub}), (\ref{h}), (\ref{frw}) and (\ref{rho}), and
the fact that $c_{\mu\nu}$ will decay rapidly outside the
anisotropy interval $\tau<t<T$, we find from (\ref{qdot}) that
\[
\begin{array}{lll}
t\ll \tau &\Rightarrow & Q\sim t \\
t\gg T &\Rightarrow & Q\sim t^{-5}
\end{array}
\]
which shows the form of quadrupole anisotropy decay outside
the anisotropy interval. By comparison, the Weyl curvature
(\ref{WCH}) decays at the same rate near the singularity, while
(\ref{h}) shows that the shear anisotropy decays more rapidly
at early times, but less rapidly at late times.

\end{document}